\documentclass[
    ,final            
  ]
  {aipproc}

\layoutstyle{6x9}

\usepackage{amssymb, amsmath}

\newcommand{\nc}{\newcommand}

\nc{\be}{\begin{equation}}
\nc{\ee}{\end{equation}}
\nc{\bea}{\begin{eqnarray}}
\nc{\eea}{\end{eqnarray}}
\nc{\nn}{\nonumber}

\nc{\markx}{$\clubsuit$}

\def\Slash#1{#1\kern-0.55em\raise.05ex\hbox{/}}
\def\slash#1{#1\kern-0.5em\raise.05ex\hbox{{$\scriptstyle /$}}}

\begin{document}    

\title{Electroweak Baryogenesis in the nMSSM
\footnote{Talk presented at SUSY06, 
the 14th International Conference on Supersymmetry and the Unification
of Fundamental Interactions, UC Irvine, California, 12-17 June 2006} 
}


\author{T. Konstandin
\footnote{e-mail: konstand@kth.se}
}{
address={
Department of Theoretical Physics, Royal Institute of Technology (KTH), 
AlbaNova University Center, 
Roslagstullsbacken 21, 106 91 Stockholm, Sweden}
}

\begin{abstract}
  In this talk, electroweak baryogenesis in the nMSSM is discussed
  following Ref~\cite{Huber:2006wf}. We focus on differences compared
  to the MSSM. We conclude that electroweak baryogenesis in the nMSSM
  is rather generic. Still, sfermions of the first two generations are
  required to be heavy to evade constraints from electric dipole
  moments.
\end{abstract}

\maketitle


\section{Introduction to Electroweak Baryogenesis and the nMSSM}

A viable baryogenesis mechanism aims to explain the observed baryon
asymmetry of the Universe (BAU), $\eta = \frac{n_B - n_{\bar B}}{s}
\approx 8.7(3) \times 10^{-11}$, and the celebrated Sakharov
conditions state the necessary ingredients for baryogenesis: (i) C and
CP violation, (ii) non-equilibrium, (iii) B number violation.

B number violation is present in the hot Universe due to sphaleron processes
while C is violated in the electroweak sector of the Standard Model (SM).
Electroweak baryogenesis (EWBG) requires a strong first-order
electroweak phase transition (PT) to drive the plasma out of equilibrium.
The CP violation is induced by the moving phase boundary and has to be
communicated into the symmetric phase, where the sphaleron process is
active~\cite{Cohen:1990it}.  Thus, the two important aspects of EWBG
are transport and CP violation. This makes it essential to derive
transport equations that contain CP-violating quantum effects in a
genuine manner.

Compared to other baryogenesis mechanisms, EWBG has the attractive
property that the relevant energy scale will be accessible by the next
generation of collider experiments.


The nMSSM of Ref.~\cite{Panagiotakopoulos:2000wp} consists of the MSSM
extended by a gauge singlet and the superpotential
\be
W_{NMSSM} = \lambda S H_1 H_2 + \frac{m^2_{12}}{\lambda} S + W_{MSSM}.
\ee
In this model, a $\mathbb{Z}_5$ or $\mathbb{Z}_7$ symmetry is imposed
to solve the domain wall problem without destabilizing the electroweak
hierarchy. The $\mu$ term is forbidden and only induced after
electroweak symmetry breaking. Thus the $\mu$ problem is solved.  The
discrete symmetries also eliminate the singlet self coupling.  A
rather large value of $\lambda$ is needed in the nMSSM to fulfill current
mass bounds on the Higgsinos and charginos, which might lead to a
Landau pole below the GUT scale.

\section{EWBG in the MSSM and its extensions}

In the MSSM and its extensions the dominant contribution to
baryogenesis comes from the charginos (Higgsino - Wino - mixing) with
the mass matrix
\bea
\psi_R = \binom{\tilde W^+_L}{\tilde h_{1,R}}, \quad 
\psi_L = \binom{\tilde W^+_R}{\tilde h_{2,L}}, \quad
m(z) = \begin{pmatrix}
M_2 & g \, H_2^* (z) \\
g \, H_1^* (z) & \mu(z)
\end{pmatrix},
\eea
where the Wino mass parameter $M_2$ and the $\mu$ parameter
contain complex phases. In the nMSSM the $\mu$ parameter is
proportional to the vev of the singlet field and hence changes during
the phase transition, while it is constant in the MSSM.

As mentioned above, to obtain unambiguous results for the predicted
BAU, a formalism is required that treats transport and CP violation in
a genuine manner. This formalism is given by the Kadanoff-Baym
equations that constitute the statistical analog to the
Schwinger-Dyson equations.

The simplest example of CP violation in transport equations is given
by the one-flavour case with a $z-$dependent complex phase in the mass
term~\cite{Joyce:1994zn}, $m(z)= |m(z)| \times e^{i \theta(z)}$. The
transport equation for the particle distribution function $f$ is in
this case of the Vlasov type ($\omega^2=k^2+m^2$)
\be
\frac{k_z}{\omega} \partial_z f_s + F_s \partial_{k_z} f_s = \textrm{collision terms}, \quad
F_s = -\frac{{|m|^2}^\prime}{2 \omega} + s \frac{(|m|^2 \theta^\prime)^\prime}
        {2 \omega \sqrt{\omega^2 - k_\parallel^2}} \label{sem_cl_force}.
\ee
Notice that the second part of
the force $F_s$ violates CP and hence sources EWBG.

\begin{figure}[t]
\centering
\includegraphics[width=0.3\textwidth]{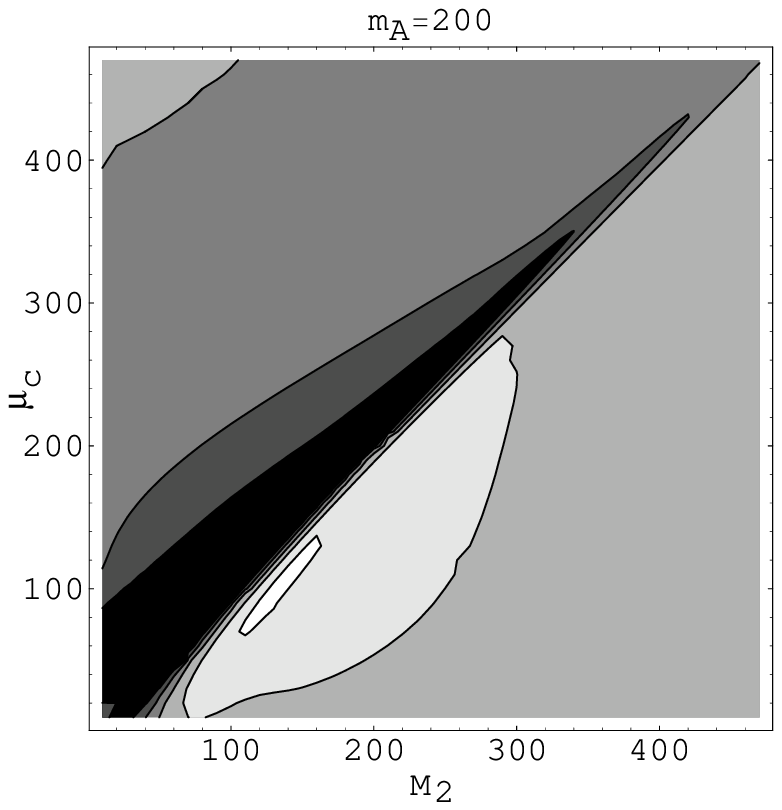}
\includegraphics[width=0.4\textwidth]{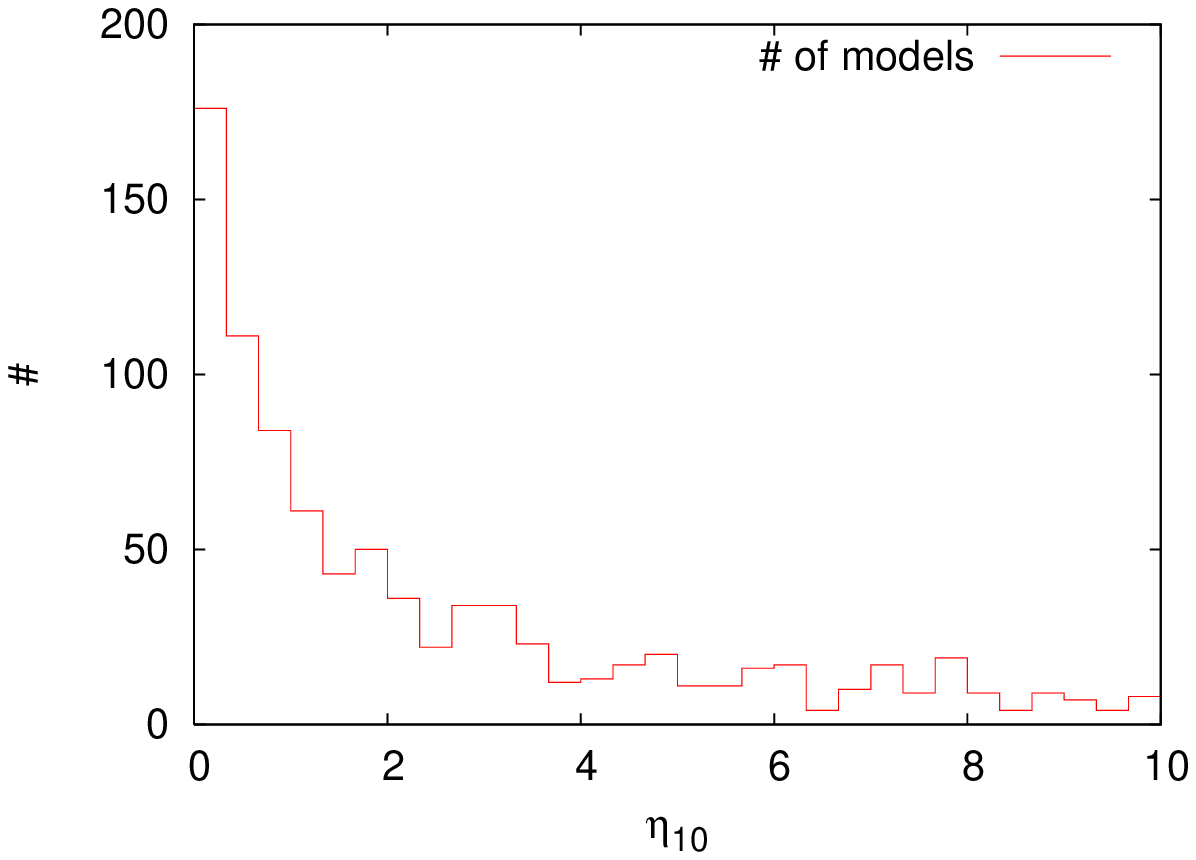}
\label{MSSM_EWBG}
\caption{
The left plot shows the produced BAU in the MSSM, $m_A=200$ GeV.
The right plot shows the produced BAU by random nMSSM models, $M_2=200$ GeV. }
\end{figure}

The multi flavour case can be treated in linear
approximation~\cite{Konstandin:2004gy}.  In this case new sources of
baryogenesis are present that are based on flavour mixing. However,
these sources are suppressed by flavour oscillations and are only
relevant for almost mass degenerate charginos. Former approaches
neglected the flavour oscillation what lead to larger results,
especially away from mass degeneracy~\cite{Carena:2000id}.  The left
plot of Fig.~\ref{MSSM_EWBG} shows the produced baryon asymmetry for a
maximal CP-violating phase in the chargino mass matrix using the
system of diffusion equations suggested in Ref.~\cite{Huet:1995sh} for
the MSSM.
The black area denotes the region of the parameter space where EWBG is
viable.  We note that EWBG in the MSSM is only possible if: (i) The
charginos are nearly mass degenerate such that mixing effects are not
suppressed.  (ii) The CP phases in the chargino sector are ${\cal O}(1)$.
Similar to chargino mediated EWBG, neutralinos can give rise to a
contribution to the BAU of similar size~\cite{Cirigliano:2006dg}. 
\subsection{Electroweak phase transition}
\begin{figure}[t]
\centering
\includegraphics[width=0.4\textwidth]{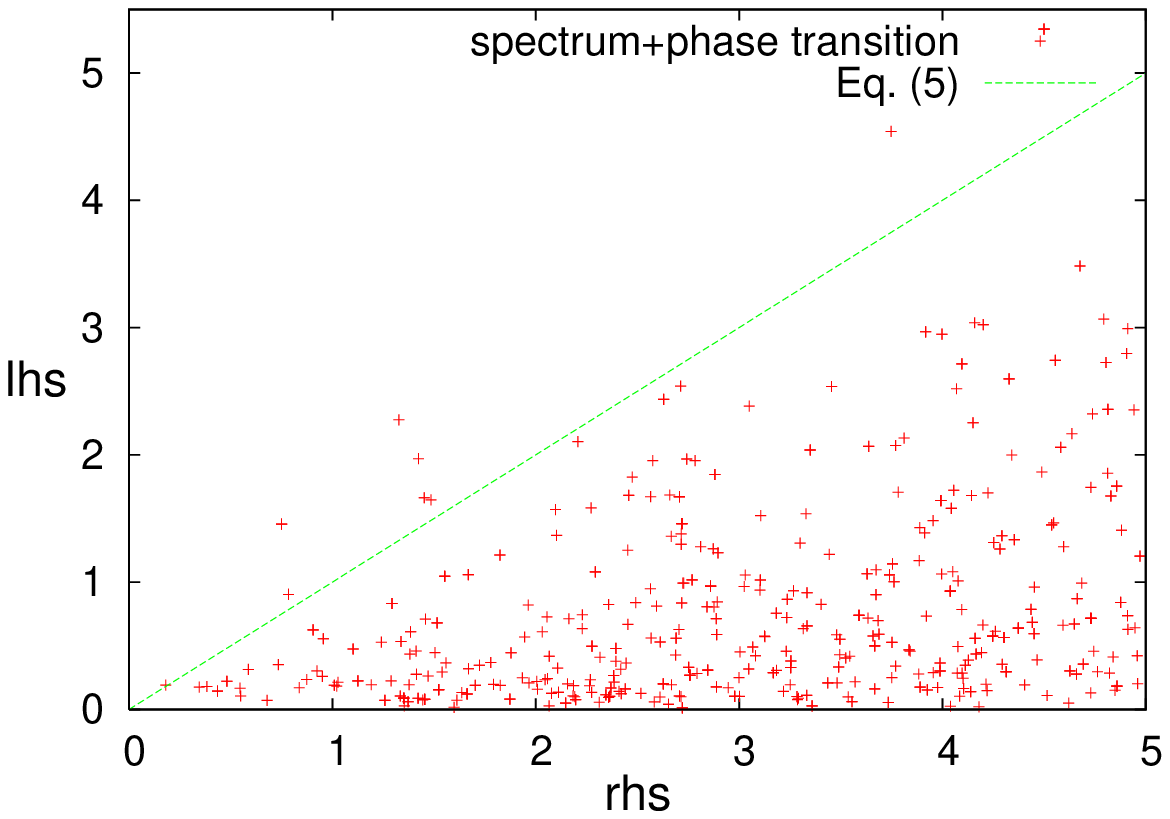}
\includegraphics[width=0.4\textwidth]{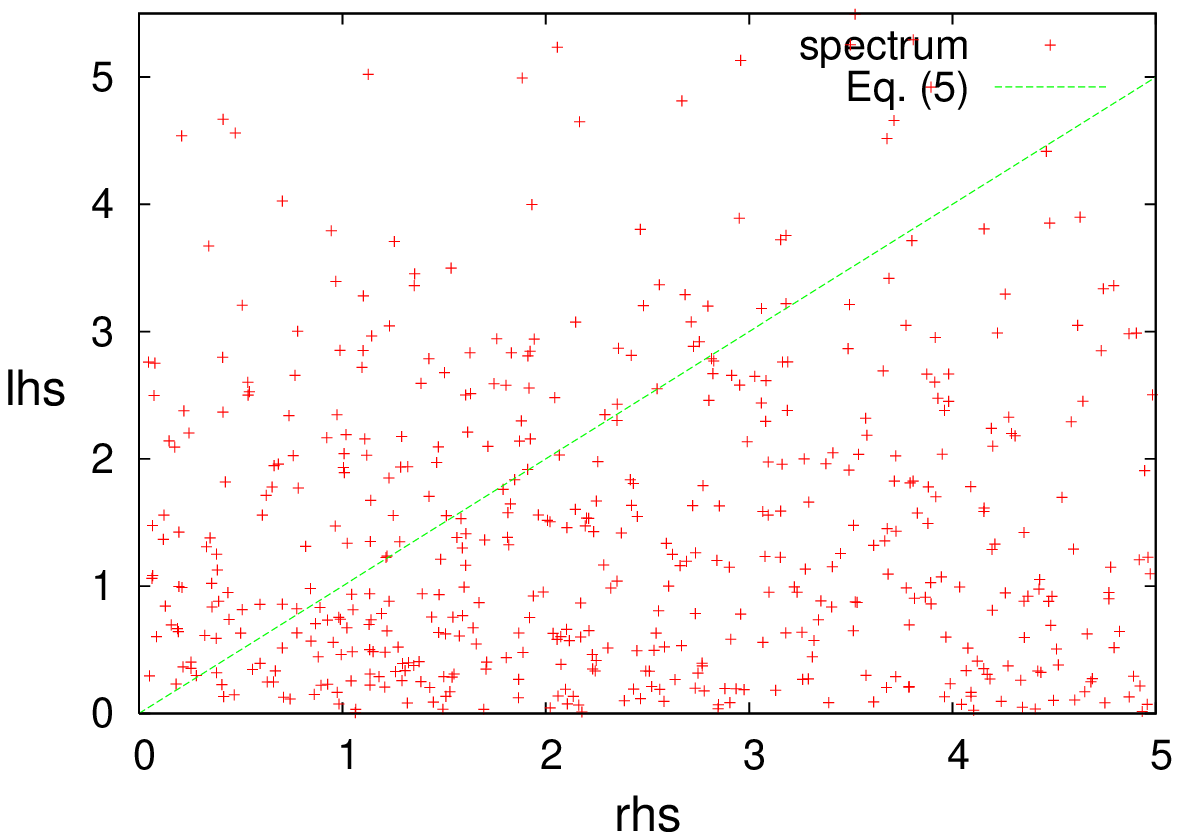}
\caption{
The left/right plot shows both sides of the inequality (\ref{PT_ineq})
for random nMSSM models with/without a first-order phase
transition.\label{PT_NMSSM}}
\end{figure}
In contrast to the MSSM, no light stop is needed in the nMSSM, since
the additional singlet terms in the Higgs potential strengthen the
phase transition~\cite{Huber:2000mg}. In the nMSSM case these terms
read:
\be
{\cal L} = {\cal L}_{MSSM} + m_s^2 |S|^2
        + \lambda^2 |S|^2 (H_1^\dagger H_1 + H_2^\dagger H_2)
 +\, t_s (S + \, h.c.) + (a_\lambda S H_1 \cdot H_2 + \, h.c.).
\ee
The parameters $a_\lambda$ and $t_s$ are SUSY breaking and all sources
of CP violation in this potential can be contributed to $t_s$. In a
simplified scheme without CP violation, a first-order phase transition
due to tree-level dynamics occurs if~\cite{Menon:2004wv}
\be
m_s^2 < \frac{1}{\tilde \lambda} \left| \frac{\lambda^2 t_s}{m_s} 
        - m_s \tilde a \right|, \quad 
\tilde a = \frac{a_\lambda}{2} \, \sin 2 \beta, \quad 
\tilde \lambda^2 = \frac{\lambda^2}{4} \sin^2 2\beta + 
\frac{\bar g^2}{8} \cos^2 2\beta.
\label{PT_ineq}
\ee
Fig. \ref{PT_NMSSM} displays Eq.~(\ref{PT_ineq}) for random nMSSM
models with/without a strong first-order PT and shows that this criterion is also
decisive if CP violation and the one-loop effective potential are
taken into account~\cite{Huber:2006wf}.

\subsection{EDM constraints and baryon asymmetry}

%
%
Since the trilinear term in the superpotential contributes to the
Higgs mass, $\tan (\beta)$ is generically of ${\cal O}(1)$. Hence two-loop
contributions from the charginos to the electron EDM are naturally
small. The one-loop contributions to the electron EDM can, as in the
MSSM, be reduced by increasing the sfermion masses.

The effective $\mu$ parameter is dynamical in the nMSSM and its
complex phase changes during the phase transition. This leads to new
CP-violating sources in the chargino sector that are of second order
in the gradient expansion as the contributions in the one flavour case
of Eq. (\ref{sem_cl_force}). These contributions do not rely on
flavour mixing and are not suppressed by the flavour
oscillations. Thus mass degenerate charginos are not required for
viable EWBG in the nMSSM.  Additionally, the bubble wall tends to be
thinner than in the MSSM what further enhances second order sources
compared to first order sources. This leads to the fact that it is
rather generic to generate the observed baryon asymmetry in the
nMSSM~\cite{Huber:2006wf}. The right plot of Fig.~\ref{MSSM_EWBG}
shows the binned BAU for a random set of nMSSM models with a strong
first order PT.

\section{Conclusions}
The nMSSM provides a framework in which electroweak baryogenesis 
seems to be possible without tuning.


\begin{theacknowledgments}
I would like to thank S. Huber, M.G. Schmidt and T. Prokopec for the
fruitful collaboration on the presented subject.  Moreover, I would
like to acknowledge financial support by the Swedish Research Council
(Vetenskapsr{\aa}det), Contract No.~621-2001-1611 and Resebidrag.

\end{theacknowledgments}



\bibliographystyle{aipproc}   


%


\end{document}